\documentclass[twocolumn,showpacs,amsmath,amssymb,showkeys,
floatfix,pra]{revtex4}
\usepackage{amsfonts,amsmath,bbold}
\usepackage{graphicx}
\usepackage{dcolumn}
\usepackage{bm}

\begin{document}

\title{Comment on ``Symplectic quantization, inequivalent quantum theories, and Heisenberg's principle of uncertainty''}

\author{D. C. Latimer}
\affiliation{Department of Physics and Astronomy\\
Valparaiso University\\ 
Valparaiso, Indiana 46383, USA}
\date{\today}

\begin{abstract}
In Phys.\ Rev.\ A {\bf 70}, 032104 (2004), M. Montesinos and G. F. Torres del Castillo consider various symplectic structures on the classical phase space of the two-dimensional isotropic harmonic oscillator.  Using Dirac's quantization condition, the authors investigate how these alternative symplectic forms affect this system's quantization.  They claim that these symplectic structures result in mutually inequivalent quantum theories.  In fact, we show here that there exists a unitary map between the two representation spaces so that the various quantizations are equivalent.
\end{abstract}

\maketitle

\section{Introduction}
In Ref.\ \cite{symp}, the authors study the effects that alternative symplectic structures have on the quantization of a two-dimensional harmonic oscillator.  Their motivation stems from the fact that a symplectic structure on the phase space of a linear system is typically chosen in terms of the canonical coordinates of the system.  For some systems, the normal modes or canonical coordinates may not be known so that this standard choice cannot be made.  These authors conclude that alternative symplectic structures lead to unitarily inequivalent quantum theories.  
This conclusion would support the notion that quantization is coordinate dependent; however, this is not the case.  

Below, we review Dirac's quantization condition with a standard symplectic form.  Then, we discuss the Fock representation of the canonical commutation relation (CCR) algebra, paying particular attention to the complexification of the (real) phase space.  It is the complex structure which selects, to a degree, the particular representation of the algebra.  Alternative symplectic forms will result in a different complexification of the phase space and, thus, an alternate quantization of the system.  The two complexified spaces are isometrically isomorphic, and we will see that the two different representations of the CCR algebra are equivalent.  Additionally, we derive the transformation that relates observables in each representation and discuss potential problems with interpretation of the quantizations.
For completeness sake, we make note of a coordinate independent approach to the quantization of linear systems found in Ref.\ \cite{kch}.  In this method, one need only know the energy and time evolution of the classical system; no explicit symplectic form is needed. 

\section{The Dirac condition}
The phase space of a classical system is real and even dimensional $V=\mathbb{R}^{2n}$.  We can define a non-degenerate symplectic form $s(\cdot,\cdot)$ on $V$.  (Note:  Montesinos and Torres del Castillo use the geometer's notation for the symplectic form $\omega^{\mu \nu}$.)  By definition, a symplectic form satisfies $s(v,w)=-s(w,v)$ for all elements in the phase space, and non-degeneracy implies that if $s(v,v)$ vanishes, then the element $v$ vanishes.
Additionally, for linear systems such as the harmonic oscillator, the time evolution for elements in phase space is determined by
\begin{equation}
\frac{\mathrm{d}v}{\mathrm{d}t} = \Omega v
\end{equation}
with $v \in V$.   Identifying vectors as $v = (x,p)$, the isotropic harmonic oscillator has
\begin{equation}
\Omega = 
\left( \begin{array}{cc}
0 & m^{-1} \\
-m \omega^2 & 0
\end{array} \right),
\end{equation}
for mass $m$ and angular frequency $\omega$, [cf.~Eq.~(11) in Ref.~\cite{symp}].
To relate these classical elements of phase $v$ to their quantum counterparts $\mathcal{Q}(v)$, Dirac suggested that the quantized operators should satisfy the commutation relation
\begin{equation}
[\mathcal{Q}(v),\mathcal{Q}(w)   ] = i \hbar s(v,w). \label{dirac}
\end{equation}
It is this condition which gives rise to the canonical commutation relations.  In a particular representation of a quantized system, physical observables form a noncommutative algebra of operators on a Hilbert space.  The situation complicates itself in that there are an infinite number of representations of such observables describing a single system.  Luckily, algebras which are unitarily equivalent turn out to describe the same physics;  only unitarily inequivalent representations describe different physical situations (for example, thermal states at different temperatures).  By using an algebraic framework, we are able to fully appreciate the subtleties involved in two different representations of a system.

\section{The CCR algebra}

Stone and von Neumann identified the conditions by which two representations were unitarily equivalent.  Their work considered a fixed symplectic structure, whereas Montesinos and Torres del Castillo deal with a host of symplectic forms \cite{symp}.  Still, the point remains that unitarily equivalent representations describe the same physics.  We shall concern ourselves with Fock representations of the CCR algebra; a concise exposition of this topic can be found in Ref.~\cite{bongaarts}, for instance.  We will carefully construct a representation of the CCR algebra, given the symplectic form 
$s(\cdot, \cdot)$.  By using a bottom-up approach, we will be able to compare the similarities and differences between the representations based upon two different symplectic forms.

The CCR algebra is constructed from creation and annihilation operators defined over a complex vector space, the one-particle space.  To this end, we need to first convert $V$ into a complex vector space with inner product, and then construct an algebra of quantized elements that satisfy the Dirac condition, Eq.~(\ref{dirac}).  
To complexify the space, we must find a positive complex structure $J$ on $V$ which is compatible with the symplectic form.  In total, we require
\begin{eqnarray}
&&J^2 = -1,  \label{minus} \\
&&s(Jv,v) \ge 0,  \label{pos}  \\
&&s(Jv,Jw) = s(v,w). \label{Jsymp}
\end{eqnarray}
The complexified space $V_J$ has complex dimension $n$ and carries the positive-definite inner product
\begin{equation}
\langle v, w \rangle_J = s(Jv, w) - i s(v,w).  \label{ip}
\end{equation}
We place an additional constraint upon the complex structure. So that time translation can be implemented by a unitary operator, the generator of time translation $\Omega$ must be complex linear
\begin{equation}
J\Omega = \Omega J. \label{gen}
\end{equation}

The Fock representation $\pi_J$ of the CCR algebra consists of operators $\pi_J(v)$, real linear in $v$, on the Hilbert space $\mathcal{H}_J$.  The elements of the algebra satisfy the commutation relation
\begin{equation}
[ \pi_J(v) ,\pi_J(w)  ]  =- i\hbar s(v,w), \label{ccr1}
\end{equation}
akin to Eq.~(\ref{dirac}).  
The Gel'fand-Naimark-Segal theorem \cite{GNS} guarantees the existence of a unique (up to phase) cyclic vector $\Phi_J$ in the Hilbert space satisfying the vacuum property
\begin{equation}
\pi_J(v + iJv) \Phi_J =0,
\end{equation}
for all $v \in V$.
In the usual treatment, these elements of the algebra are called annihilators.  For ease of interpretation, it is useful to define the annihilation operators
\begin{equation}
a_J(v) =\frac{1}{\sqrt{2}} \pi_J(v + iJv);
\end{equation}
creation operators are the adjoints of the annihilators, $a_J(v)^*= \frac{1}{\sqrt{2}} \pi_J(v - iJv)$ .  From the real linearity of $\pi_J(\cdot)$, one can show that the creators are complex linear operators over $V_J$. Additionally, the commutation relation in Eq.~(\ref{ccr1}) results in the familiar expression of the CCR
\begin{equation}
[a_J(v),  a_J(w)^*   ]  = \hbar \langle v, w  \rangle_J, \qquad
[a_J(v),  a_J(w)   ]  = 0.
\end{equation}

In any particular representation, self-ajdoint operators that are of Hamiltonian type (e.g., second-quantised Hamiltonians, number operators, and the like) satisfy the commutation relation
\begin{equation}
[\widetilde{H}_J, a_J(v)^*] = \hbar\, a_J(H_Jv)^* \label{ham_cr}
\end{equation}
where $\widetilde{H}_J$ is the second-quantised operator corresponding to the single particle operator $H_J$ on $V_J$.   In addition to this relation, we renormalize the vacuum so that such operators annihilate the vacuum, $\widetilde{H}_J \Phi_J =0$.
Operators in field theory can be expressed in terms of the creators and annihilators. Using the CCR, one can show that the above operator can be written as
\begin{equation}
\widetilde{H}_J = \sum_{j,k} \langle e_j, H_J e_k \rangle_J\, a_J( e_j)^* a_J(e_k) \label{ham_rep}
\end{equation}
where $\{e_j\}$ form an orthonormal basis of $V_J$.
Expressing these operators in terms of the creators/annihilators allows us to determine how such operators transform between representations.

Reflecting on the r\^ole of the symplectic form, we note that it is involved in determining the complex structure and inner product on $V_J$.  Furthermore, the complex structure determines the form of the annihilators/creators and vacuum vector.
Clearly, a change in the complex structure results in a different representation.  
For a fixed symplectic form, the effect of alternative complex structures has been extensively studied.  It is well known that representations based on such alternative complex structures are related via Bogoliubov transformations.  In what follows,  we will attempt to understand the ramifications of choosing alternative symplectic forms.  We will find that alternative forms yield alternative complex structures which are relatable to the standard one via invertible linear transformations.

\section{Alternative symplectic forms} \label{asf}

Using the same rubric as above, we may construct another representation of the CCR algebra using an alternative symplectic form $s'(\cdot,\cdot)$.  
By maintaining abstract notation, this conclusion will be valid for all of the alternative forms considered in Ref.~\cite{symp}.  What is more, our results are valid for general linear systems, not just the harmonic oscillator.

Let  $s'(\cdot, \cdot)$  be another non-degenerate symplectic form on $V$.  Given this new form, we must ``forget" the previous rules developed above that defined the complex inner product space. 
Instead, we find another positive complex structure $K$ compatible with the new symplectic form.  The inner product on $V_K$ can be got in an analogous manner to Eq.\ (\ref{ip}).  From this, we may construct another Fock representation $\pi_K$ of the CCR algebra with vacuum vector $\Phi_K$.  The creators and annihilators built from $\pi_K$ will satisfy the analogous CCR
\begin{equation}
[a_K(v), a_K(w)^*]= \hbar \langle v, w \rangle_K , \qquad
[a_K(v),  a_K(w)   ]  = 0.
\end{equation}
It is important to appreciate that this inner product space is different from that of $V_J$; below we will explore some of the oddities which emphasize this difference.   Subtle differences in the complexifications  carry through to the representations and Hilbert spaces.  Attempting to determine the action of an operator in one  representation $\pi_K$ on the other Hilbert space $\mathcal{H}_J$ can produce a nonsensical result.

A relationship can be found between the two constructed representations by noting that any two non-degenerate symplectic forms on a finite dimensional space are equivalent.  There exists an invertible linear transformation $g$ on $V$ such that 
\begin{equation}
s'(v,w) = s(g^{-1}v, g^{-1} w) \label{equiv}
\end{equation}
for all elements $v,w \in V$.  This is easily seen by recalling that for each symplectic form one may construct a symplectic frame; the transformation $g$ is then just the map between frames.  
This transformation is by no means unique.  For instance, composition of $g$ with a transformation $h$ that preserves the symplectic form $s(\cdot, \cdot)$ also relates the two forms
\begin{equation}
s'(v,w) = s(h^{-1}g^{-1}v,h^{-1}g^{-1} w) = s(g^{-1} v, g^{-1} w).
\end{equation}
Nevertheless, we need only know of the transformation's existence.

We now jointly consider the two complexifications of $V$.
Given the positive complex structure $J$ compatible with $s(\cdot, \cdot)$, the operator $K= g J g^{-1}$ is a positive complex structure compatible with $s'(\cdot, \cdot)$.  The proof of this is trivial.  Given this, the map $g: V_J \to V_K$ is an isometric isomorphism.  Isometry can be seen easily by applying the definitions of the inner products so that
\begin{equation}
\langle v, w \rangle_J  = \langle gv, gw \rangle_K. \label{isometry}
\end{equation}
As an aside, we note that if the alternative symplectic form is identical to the original form then the transformation $g$ in Eq.\ (\ref{equiv}) is a symplectic transformation.  This would lead to Bogoliubov transformations of the Fock representation.  As our motive is to investigate alternative symplectic forms, we are specifically interested in invertible transformations which are {\it not} symplectic.

The question remains as to the relation between the two different Fock representations of the CCR algebra.  We recall that an $n$--particle vector in Fock space can be identified with an element of the $n$--fold symmetric tensor space $\bigotimes_S^n V$.  In fact, with the standard inner product on the tensor space, the span of such elements in Fock space are isomorphic to the tensor space.  With this, it is clear that the map $g: V_J \to V_K$ then induces a unitary isomorphism from $\bigotimes_S^n V_J$ to $\bigotimes_S^n V_K$ for each $n$ via the action
\begin{equation}
v_1 \otimes_S \cdots \otimes_S v_n \mapsto gv_1 \otimes_S \cdots \otimes_S gv_n.  
\end{equation}
This extends to a unitary isomorphism between the two Fock spaces $U_g: \mathcal{H}_J \to \mathcal{H}_K$  with
\begin{equation}
U_g \pi_J(v) U_g^* = \pi_K(gv), \qquad U_g \Phi_J = \Phi_K.
\end{equation}
Using this fact, one may see that annihilators (creators) map to annihilators (creators)
\begin{equation}
U_g a_J (v) U_g^* = a_K( g v). \label{a_map}
\end{equation}
It is a simple exercise to show that the two-point correlation functions in both representations agree
\begin{equation}
\langle \Phi_J, \pi_J(v)^* \pi_J(w) \Phi_J \rangle_J = \langle \Phi_K, \pi_K(gv)^* \pi_K(gw) \Phi_K \rangle_K.
\end{equation}
As Fock states are quasifree,  all correlations are determined by the one- and two-point correlation functions.  Given this, correlations in both representations will agree.
From this unitary equivalence, we see that the physical content of the theory does not depend upon the particular representation chosen, and thus the physics is independent of the choice of symplectic structure. 

As  a point of contrast, we review the requirements for the implementability of Bogoliubov transformations.  Here, one deals with transformations $h$ which preserve the symplectic form.  The central question is to determine what conditions admit an implementable unitary transformation $U_h: \mathcal{H}_J \to \mathcal{H}_J$ such that
\begin{equation}
U_h \pi_J(v) U_h^* = \pi_J (hv).
\end{equation}
It is known that this this transformation is implementable whenever the operator $hJ-Jh$ is Hilbert-Schmidt; this requirement demands that the conjugate linear portion of $h$ not be ``too large" in the Hilbert-Schmidt sense.  A general transformation maps annihilators to a sum of a creator and annihilator, unlike the situation shown in Eq.~(\ref{a_map}).  If such transformations are not implementable, then one may have unitarily inequivalent representations of the CCR algebra.  

Returning to our immediate considerations, we may use the map in Eq.~(\ref{a_map}) to determine how second quantized operators transform.  For those of Hamiltonian type as in Eq.~(\ref{ham_cr}), we have 
\begin{equation}
U_g \widetilde{H}_J U_g^* = \sum_{j,k}  \langle e_j, H_J e_k \rangle_J\, a_K( g e_j)^* a_K( g e_k).
\end{equation}
Using the isometry between the spaces $V_J$ and $V_K$ and Eq.~(\ref{isometry}), we may write the inner products on the RHS as
\begin{equation}
\langle e_j, H_J e_k \rangle_J = \langle g e_j,  H_K g e_k \rangle_K,
\end{equation}
where the representation of the one-particle Hamiltonian on $V_K$ is given by $H_K = g H_J g^{-1}$.  Setting $f_j = ge_j$, then isometry implies that the set $\{ f_j \}$ is orthonormal on $V_K$.  With this recognition, the representation of the operator on $\mathcal{H}_K$ follows the same form as Eq.~(\ref{ham_rep})
\begin{equation}
\widetilde{H}_K = \sum_{j,k} \langle f_j, H_K f_k \rangle_K\,  a_K ( f_j)^* a_K(f_k).
\end{equation}
This exercise demonstrates the correct representation of Hamiltonian-like operators whenever alternative symplectic structures are employed.  The behavior of the second-quantized operator is governed by that of the single-particle operator for these two representations.  For the two representations considered here, we see that the essential features of the Hamiltonians remain the same.  The expansions in terms of the creators/annihilators are done in different bases; however, the eigenvalues of the operators must be the same.  The only change in the one-particle operator is conjugation by an invertible transformation; this does not affect the eigenvalues. In particular, if $H_J$ is a positive operator, then $H_K$ is also positive, likewise for the second-quantized operators.  Despite the choice of symplectic structure, the action of the Hamiltonian is unchanged.

\section{A simple example} 

Problems can arise if one carelessly mixes about elements of the two different representations.  We will consider a simple system which will be useful in elucidating some of the pitfalls encountered in Ref.~\cite{symp}.   To set the stage, we rescale the coordinates of the isotropic harmonic oscillator over $V=\mathbb{R}^{4}$.  Our simple system satisfies the linear equation
\begin{equation}
\partial_t v =  \left( \begin{array}{cc}
0 & \omega \\
- \omega & 0
\end{array} \right) v. \label{ham_simp}
\end{equation}
If one employs the standard symplectic form $s(v,v') = p_x' x + p'_y y -  p_x x' - p_y y'$, then a positive complex structure compatible with this symplectic form is 
\begin{equation}
J= \left( \begin{array}{cc}
0 & I \\
-I & 0
\end{array} \right),
\end{equation}
where $I$ is a two-dimensional identity matrix.
The evolution equation for the classical system is then $\partial_t v = \omega J v$.  We may consider the space as a complex inner product space as outlined above, $V_J = \mathbb{C}^2$.   Through the complex structure, an element $\alpha$ in one copy of $\mathbb{C}$ can be identified with the real space via $\Re(\alpha)=p_x$ and $\Im(\alpha) = x$.   An element $\beta$ in the other copy can be identified with the $y$ modes of oscillation, $\Re (\beta) = p_y$ and $\Im(\beta) = y$.
The one particle Hamiltonian of the complexified oscillator is then trivial $H_J =  \omega I$.  The inner product on this space is as expected, and the Hamiltonian is clearly positive.

One advantage of the algebraic formulation of the associated field theory is the ease with which the spectrum of the Hamiltonian can be determined.   The vacuum vector $\Phi_J$ in the Fock representation is cyclic.  Given an orthonormal basis of the one-particle space $\{ e_j \}$, any element of the Fock space can be expressed as a linear combination of elements of the form $a_J(e_1)^{*n_1} a_J(e_2)^{*n_2} \Phi$; vectors of this sort actually form an orthonormal basis for the representation space.  Let us focus on one particular such element $\Psi $ and assume that the energy of this element is $E$; that is, $\widetilde{H}_J \Psi = E \Psi$. For the isotropic oscillator, the one-particle Hamiltonian evaluated on eigenvectors results in $H_Je_j =  \omega e_j$.  
Given this, what is the energy of the vector $a_J(e_j)^* \Psi$?  
Using the commutation relation in Eq.~(\ref{ham_cr}), we have
\begin{eqnarray}
\widetilde{H}_J a_J(e_j)^* \Psi &=& [\hbar\, a_J(H_J e_j)^* + a_J(e_j)^* \widetilde{H}_J] \Psi\\
&=& [\hbar \omega + E] a_J (e_j)^* \Psi.
\end{eqnarray}
From this result, one may prove via induction the energy of the element $\Psi =  a_J(e_1)^{*n_1} a_J(e_2)^{*n_2} \Phi_J$ in Fock space
\begin{equation}
\widetilde{H}_J \Psi =  \hbar \omega [n_1 + n_2] \Psi.
\end{equation}
As such vectors form an orthonormal basis for the Fock space, one finds the expected spectrum of the Hamiltonian.

We can quantize the system using a slightly modified symplectic form.  We set $s'(v,v') = -p_x' x + p'_y y +  p_x x' - p_y y'$; this is the second alternative form considered in Ref.~\cite{symp}.  The related complex structure for this choice is \begin{equation}
K= \left( \begin{array}{cc}
0 & I' \\
-I' & 0
\end{array} \right)
\end{equation}
where $I'$ is the two-dimensional matrix
\begin{equation}
I'= \left( \begin{array}{cc}
-1 & 0 \\
0 & 1
\end{array} \right). \label{iprime}
\end{equation}
The following transformation relates this framework to the standard symplectic form
\begin{equation}
g =\left( \begin{array}{cc}
I' & 0 \\
 0 & I
\end{array}  \right).
\end{equation}
Effectively, this maps $i \mapsto -i$ for the copy of the complexified parameter space associated with the $x$ modes of oscillation.
Returning to the classical evolution of the system in Eq.~(\ref{ham_simp}), one may write this as
\begin{equation}
\partial_t v = K \left( \begin{array}{cc}
\omega I' & 0 \\
0 & \omega I'
\end{array} \right) v.
\end{equation}
The Hamiltonian in this representation appears to be $ \omega I'$.  If this were the case, then the Hamiltonian would have both positive and negative eigenvalues; that is, in this representation, it does not appear to be bounded from below.  Recalling the transformation of operators from the previous section, we see that the one particle Hamiltonian in this representation should be $H_K = g H_J g^{-1} =  \omega I$.  What is the source of discrepancy?

The evolution of the system in the real phase space is only part of the consideration.  The key to the discrepancy is the change in complex structure associated with the different representations.  The different complex complex structure alters the identification of $V_K = \mathbb{C}^2$ with the real vector space.  Now, an element $\alpha$ in one copy of $\mathbb{C}$ is related to the real phase space via $\Re(\alpha)=p_x$ and $\Im(\alpha) = -x$.   The identification for the $y$ mode of oscillation is unchanged.  Given this, the evolution of elements of the complex space can be got from examining the evolution of real elements of the form $(-x,y,p_x,p_y)$; this vector is just $gv$.  Making this change, one finds the evolution of such elements in phase space to be $\partial_t gv = \omega K  gv$; this results in a one particle Hamiltonian $H_K = \omega I$ on the complexified space.  

Determining the spectrum of the second-quantised Hamiltonian $\widetilde{H}_K$ follows the same procedure as outlined above.  The essential features of the proof are the commutation relations satisfied by the operator, Eq.~(\ref{ham_cr}), and the action of the one-particle Hamiltonian.  If one uses as a basis the normal modes $f_j = ge_j$, then the action is simple $H_K f_j=  \omega f_j$.  The determination of the spectrum of the Hamiltonian $\widetilde{H}_K$ will exactly mirror the above arguments resulting in the expected spectrum $\hbar \omega (n_1 + n_2)$.

We are not required to expand the second-quantized operators in terms of 
the basis $\{ f_j\}$; we may use any orthonormal basis.  However, in a different basis the identification between elements in the complex and  real spaces can become confused. As demonstrated, misidentification can lead one to incorrectly conclude that the Hamiltonian in the representation based upon the alternative symplectic form is not bounded from below.  This was one of the errors encountered by Montesinos and Torres del Castillo in Ref.~\cite{symp}.  From our simple model, we demonstrate the subtleties involved in determining the actions of second-quantized operators, and ultimately we show the equivalence of the two representations.

\section{The harmonic oscillator}

The previous simple example applied the algebraic results developed for alternative symplectic forms in Section \ref{asf}.  From this example, one may fully appreciate the importance of the comlexification scheme and its tie to the choice of symplectic form.  This particular example was motivated by the second alternative form in Ref.~\cite{symp}.  We could study in detail each of the other two alternative forms considered in Ref.~\cite{symp}; however, this would be repetitive given the general algebraic proofs contained herein. Instead, for each of the alternative symplectic forms $s_j(\cdot, \cdot)$, we will determine a linear transformation $g_j$ which relates this form to the standard one as in Eq.\ (\ref{equiv}); we will also determine the complex structure $J_j$ compatible with each alternative form. (The index $j$ follows the labeling scheme in Ref.\ \cite{symp}.)  The quantized oscillators developed using these different forms will all be equivalent, and the Hamiltonians in each theory will have the same spectrum if one carefully complexifies the real phase space.  

0) We begin with the standard symplectic form in Ref.~\cite{symp}; it is given by
\begin{equation}
s_0(v, v') = p_x' x  + p_y' y - p_x x' -p_y y'. \label{s1}
\end{equation}
To determine the associated complex structure, we begin with the requirement in Eq.\ (\ref{gen}); an operator which commutes with the generator of time translation has the block form
\begin{equation}
\left( \begin{array}{cc}  
A & B \\
-m^2 \omega^2 B & A
\end{array}  \right).
\end{equation}
An operator of this form, which also satisfies Eqs.\ (\ref{minus})--(\ref{Jsymp}), yields a complex structure
\begin{equation}
J_0 = 
\left( \begin{array}{cc}  
0 & (m \omega)^{-1} \\
-m \omega  & 0
\end{array}  \right). \label{j1}
\end{equation}
Considering elements of the form $v + iJ_0 v$, annihilators in the coordinate representation are familiar
\begin{equation}
x + \frac{\hbar}{m\omega}\frac{\partial}{\partial x}, \qquad  y + \frac{\hbar}{m\omega}\frac{\partial}{\partial y}.
\end{equation}
These operators will annihilate the vacuum vector
\begin{equation}
\Phi_0 = \sqrt{m \omega/\pi \hbar}  \exp [ -m\omega( x^2 + y^2)/2 \hbar  ]. \label{vac1}
\end{equation}

i) The first alternative symplectic form considered is 
\begin{equation}
s_1(v, v') = p_x' y + p_y' x - p_x y' - p_y x'.
\end{equation}
In block form, the complex structure satisfying Eqs.\ (\ref{minus})--(\ref{Jsymp}) and (\ref{gen}) is
\begin{equation}
J_1 = \left( \begin{array}{cc}
0 & (m\omega)^{-1} G \\
 -m \omega G &  0
\end{array}  \right),
\end{equation}
where for shorthand we define the $2\times 2$ matrix
\begin{equation}
G =\left( \begin{array}{cc}
0 & 1 \\
 1 &  0
\end{array}  \right).
\end{equation}
The following transformation relates this new symplectic form to the standard one,
\begin{equation}
g_1 =\left( \begin{array}{cc}
I & 0 \\
 0 & G
\end{array}  \right);
\end{equation}
additionally, we see that $J_1 = g_1 J_0 g_1^{-1}$.   This transformation swaps the two components of momenta as was seen in this representation of the oscillator in Ref.\ \cite{symp}.

ii) The second alternative symplectic form considered is 
\begin{equation}
s_2(v, v') = -p_x' x  + p_y' y + p_x x' - p_y y'.
\end{equation}
The associated complex structure is
\begin{equation}
J_2 = \left( \begin{array}{cc}
0 & (m\omega)^{-1} I' \\
 -m \omega I' &  0
\end{array}  \right),
\end{equation}
with $I'$ defined in Eq.\ (\ref{iprime}).
The following transformation relates this framework to the standard one
\begin{equation}
g_2 =\left( \begin{array}{cc}
I & 0 \\
 0 & I'
\end{array}  \right).
\end{equation}

iii) The final alternative symplectic structure considered is 
\begin{equation}
s_3(v, v') = m\omega(p_x' p_y - p_y' p_x)+  (m\omega)^{-1} (x'y-y'x) . \label{s4}
\end{equation}
This example emphasizes the point that the alternative quantizations are related 
via general linear transformations.  Using the standard quantization procedure, dimensional analysis shows that the symplectic form should take on values whose dimensions are that of Planck's constant.  Taken at face value, it is clear that the RHS of the  symplectic form in Eq.\ (\ref{s4}) does not have these dimensions.  In fact, the terms involving momenta do not even have the same dimensions as the terms involving position.   The reason for this will become transparent whenever we determine the linear transformation relating this symplectic form to the standard one. 
The complex structure associated with this form is
\begin{equation}
J_3= \left( \begin{array} {cccc}
0 & -(m \omega)^{-2} & 0 & 0\\
(m \omega)^{2} & 0 & 0 & 0\\
0 & 0 & 0 & -(m \omega)^{-2} \\
0 & 0 & (m \omega)^{2} & 0 
\end{array} \right),
\end{equation}
and a transformation relating $J_3$ to $J_0$ is
\begin{equation}
g_3 = \left( \begin{array} {cccc}
1 & 0 & 0 & 0\\
0 & 0 & -m\omega & 0\\
0 & -(m \omega)^{-1} & 0 & 0 \\
0 & 0 & 0 & 1 
\end{array} \right).
\end{equation}
From this transformation, we see that the $x$-coordinate of momentum has been mapped to the $y$-coordinate of position, and vice versa.  Given this, along with the appropriate multiplicative factors from $g_3$, the symplectic form $s_3(\cdot, \cdot)$ does turn out to have the dimensions of Planck's constant.

In conclusion, M. Montesinos and G.F. Torres del Castillo explored the effect of alternative symplectic structures upon the quantization of a classical system \cite{symp}.   Their work is interesting in that one often takes for granted knowledge of the canonical coordinates; however, their conclusion that alternative symplectic structures lead to inequivalent quantum theories is incorrect. We noted that non-degenerate symplectic forms are related via various invertible transformations.  Given this fact, we show that quantizations based upon two different forms are relatable via a unitary map between the two representation spaces.  As a result, correlation functions in the two representations agree so that the physical content of the theory, including Heisenberg's uncertainty principle, cannot be altered by the choice of representation and, thus, symplectic structure. Through a specific example, we show that alternative symplectic forms force a change in the way in which the real phase space is complexified.  This change results in a different identification between the comlex one-particle space and the real phase space.  Ignoring this, it appears that the eigenvalues of operators, such as the Hamiltonian, change with the alternative form; however, if the subtleties of the alternative complexification are fully appreciated, it is clear that the theories are in fact equivalent.

\bibliography{symplectic}
\end{document}